\documentclass[fleqn,usenatbib]{mnras}

\usepackage{newtxtext,newtxmath}

\usepackage[T1]{fontenc}

\DeclareRobustCommand{\VAN}[3]{#2}
\let\VANthebibliography\thebibliography
\def\thebibliography{\DeclareRobustCommand{\VAN}[3]{##3}\VANthebibliography}

\usepackage{graphicx}	
\usepackage{amsmath}
\usepackage{hyperref}

\title[Mass gap black holes from stellar collisions]{Formation of black holes in the pair-instability mass gap: Hydrodynamical simulations of a head-on massive star collision}
\author[A. Ballone et al.]{
Alessandro Ballone,$^{1,2,3}$\thanks{E-mail: ale.ballone@gmail.com} 
Guglielmo Costa,$^{1,2,3}$  
Michela Mapelli,$^{1,2,3}$\thanks{E-mail:michela.mapelli@unipd.it} 
Morgan MacLeod,$^{4}$  \newauthor
Stefano Torniamenti,$^{1,2,3}$
and Juan Manuel Pacheco-Arias$^1$\\
$^{1}$Physics and Astronomy Department Galileo Galilei, University of Padova, Vicolo dell'Osservatorio 3, I-35122 Padova, Italy\\
$^{2}$INAF - Osservatorio Astronomico di Padova, Vicolo dell'Osservatorio 5, I-35122 Padova, Italy\\
$^{3}$INFN - Padova, Via Marzolo 8, I--35131 Padova, Italy\\
$^{4}$Center for Astrophysics $\vert$ Harvard $\&$ Smithsonian 
60 Garden Street, MS-16, Cambridge, MA 02138, USA\\
}

\date{Accepted XXX. Received YYY; in original form ZZZ}

\pubyear{2022}

\begin{document}
\label{firstpage}
\pagerange{\pageref{firstpage}--\pageref{lastpage}}
\maketitle

\begin{abstract}
The detection of the binary black hole merger GW190521, with primary black hole mass $85^{+21}_{-14}$ ${\rm M}_{\odot}$, proved the existence of black holes in the theoretically predicted pair-instability gap ($\sim60-120 \, {\rm M}_{\odot}$) of their mass spectrum. Some recent studies suggest that such massive black holes could be produced by the collision of an evolved star with a carbon-oxygen core and a main sequence star. Such a post-coalescence star could end its life avoiding the pair-instability regime and with a direct collapse of its very massive envelope. It is still not clear, however, how the collision shapes the structure of the newly produced star and how much mass is actually lost in the impact. We investigated this issue by means of hydrodynamical simulations with the smoothed particle hydrodynamics code {\sc StarSmasher}, finding that a head-on collision can remove up to 12\% of the initial mass of the colliding stars. This is a non-negligible percentage of the initial mass and could affect the further evolution of the stellar remnant, particularly in terms of the final mass of a possibly forming black hole.
We also found that the main sequence star can plunge down to the outer boundary of the core of the primary, changing the inner chemical composition of the remnant. The collision expels the outer layers of the primary, leaving a remnant with an helium-enriched envelope (reaching He fractions of about 0.4 at the surface). These more complex abundance profiles can be directly used in stellar evolution simulations of the collision product. 
\end{abstract}




\section{Introduction}\label{intro}

Stellar evolution models predict a gap in the mass spectrum of black holes between $\sim60$ and $\sim120\,{}{\rm M}_{\odot}$ \citep[the so-called ``pair instability mass gap''; e.g.,][]{Heger02, Woosley07,Belcz16,Spera17, Woosley17, Stevenson19,Marchant19,Farmer19, Leung19, Marchant19,Vigna-Gomez19, Marchant20, Renzo20a, Tanikawa21, Mehta22, Rahman22}. Pair instability affects those massive stars that develop carbon-oxygen (CO) cores with densities between $\approx10^2-10^6  \mathrm{\;g \;cm^{-3}}$ and temperatures above $6\times 10^8$~K. For these physical conditions, thermal energy is converted into mass of electron-positron pairs. Depending on the CO core mass, this sudden lack of thermal support can lead to cyclic phases of hydrodynamical instability of the star that can result into strong mass loss (pulsational pair-instability) or even to a single hydrodynamical instability phase that leads to the total collapse of the star and completely destroys it (pair-instability supernova).

Gravitational-wave observations seem to challenge our theoretical understanding of such gap. During the third observing run, the LIGO--Virgo collaboration detected a black hole merger event, GW190521, with primary and secondary mass $M_1=85^{+21}_{-14}$ and $M_2=66^{+17}_{-18}\,{} {\rm M}_{\odot}$, respectively
 \citep{Abbott20a,Abbott20b}. The mass of the primary black hole of GW190521 lies within the predicted mass gap, while the mass of the secondary is close to its lower boundary. Furthermore, three additional gravitational-wave event candidates might be associated with black holes in the mass gap \citep{abbottGWTC2.1, abbottGWTC3}: GW190403\_051519 ($M_1=85^{+28}_{-33} {\rm M}_{\odot}$, $M_2=20^{+26}_{-8}{\rm M}_{\odot}$), GW190426\_190642 ($M_{1}=106^{+45}_{-24} {\rm M}_{\odot}$, $M_{2}=76^{+26}_{-36}{\rm M}_{\odot}$) and GW200220\_061928 ($M_{1}=87^{+40}_{-23} {\rm M}_{\odot}$, $M_{2}=61^{+26}_{-25}{\rm M}_{\odot}$). Finally, \cite{Nitz21} recently reported another potential event candidate with primary mass overlapping with the mass gap (GW200129\_114245; $M_{1}=79^{+40}_{-38} {\rm M}_{\odot}$, $M_{2}=32^{+19}_{-14}{\rm M}_{\odot}$).

 Our knowledge of the boundaries of the pair-instability mass gap is hampered by several uncertainties about massive stellar evolution: recent work has shown that the gap might be substantially shorter than initially predicted \citep[e.g.,][]{Croon20, Farmer20, Costa21, Vink21, Woosley21, Siegel22}.
 Alternatively, GW190521 might be a 2nd generation black hole merger, in which the primary and maybe also the secondary black hole are the result of a previous merger. Such hierarchical mergers are shown to be occurring in star clusters \citep[][]{Miller02,Gerosa17,Fishbach17,Rodriguez19,Antonini19,Arca-Sedda20,Fragione20, Rodriguez20,Arca-Sedda21b,Arca-Sedda21,Mapelli21a,Mapelli21b, Mandel22, vynatheya2022, Rizzuto22} or triggered by massive gaseous discs in active galactic nuclei \citep[e.g.,][]{McKernan12,McKernan18,Bartos17,Stone17,Yang19,Tagawa20,Tagawa21}.

 Finally, stellar collisions provide an additional pathway for the formation of black holes in the pair-instability gap  \citep{DiCarlo19,DiCarlo20b,DiCarlo20,Spera19,Kremer20, Rastello21, Rizzuto21, Banerjee22}. 
 Simulations of young massive star clusters show that dynamical encounters efficiently trigger collisions between massive stars. If an evolved star with a relatively low-mass core collides with a massive star in its main sequence (MS), the product of such collision could consist of a star with the same core, but with a much more massive envelope. 
 This exotic star can, in principle, avoid the pair-instability phase and eventually evolve until it directly collapses into a black hole with mass comparable to the sum of the masses of the two colliding stars \citep{DiCarlo19,DiCarlo20b,DiCarlo20}. 
 
 \citet{Renzo20b} focused on a specific stellar collision described in \citet{DiCarlo20}, involving a core helium burning star of $\approx58$~M$_{\odot}$ 
 and a MS star of $\approx42$~M$_{\odot}$.  
 In their study, they used the stellar evolution code {\sc mesa} \citep[][]{Paxton11,Paxton19} to compute the stellar structures of both stars for a time equal to the end of the MS of the most massive star. Then, they relaxed the primary star by adding the mass of the secondary to its envelope (hence reaching a total mass of $\approx 100$~M$_{\odot}$) and assuming that the envelope of the primary is enriched - uniformly with radius - with He brought by the H-burning core of the secondary. In this way, \citet{Renzo20b} found that this system can directly collapse into a black hole in the pair-instability mass gap.
 
 One of the main uncertainties of the approach of \citet{DiCarlo20} and \citet{Renzo20b} is in how the collision actually shapes the collision product, both in terms of mass loss and in terms of chemical enrichment of the primary's envelope. The only way to estimate the mass lost during the collision and to assess the level of chemical mixing in the merger product is to perform a three-dimensional hydrodynamical simulation of the collision. While a certain number of hydrodynamical simulations of stellar collisions have been carried out \citep[for head-on or low impact-parameter collisions, see ][]{Lombardi96, Sills01, Lombardi02,Gaburov10,Glebbeek13}, very few involve extremely massive stars in their post-MS phase, as those possibly leading to black holes in the pair-instability gap. 
 
 In this work, we simulate the collision of two massive stars (similar to the ones considered by \citealt{DiCarlo20} and \citealt{Renzo20b}), and study the structure of the collision product, by means three-dimensional hydrodynamical simulations. We find that up to $\sim{12}\%$ of the initial mass is lost during the collision, and quantify the chemical mixing in the post-coalescence product.

\begin{figure}
\includegraphics[scale=0.62, trim={0.8cm 0 0cm 0},clip]{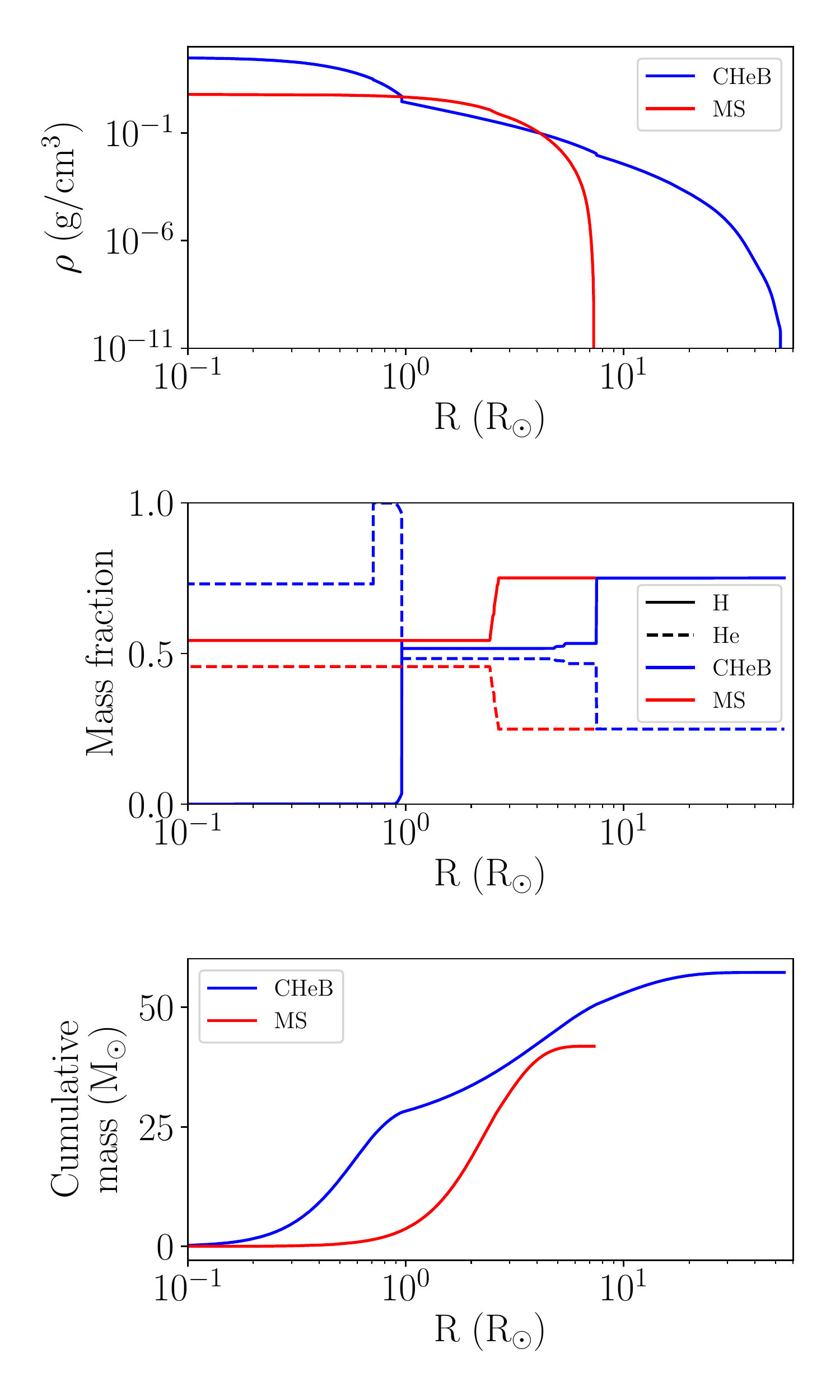}
\caption{Density (upper panel), hydrogen and helium mass fraction (middle panel: H, solid line; He, dashed line) and cumulative mass profiles (lower panel) for the CHeB (blue) and MS star (red), for our initial conditions.\label{ics}}
\end{figure}

\begin{figure*}
\includegraphics[scale=0.305, trim={1.3cm 0 1cm 0},clip]{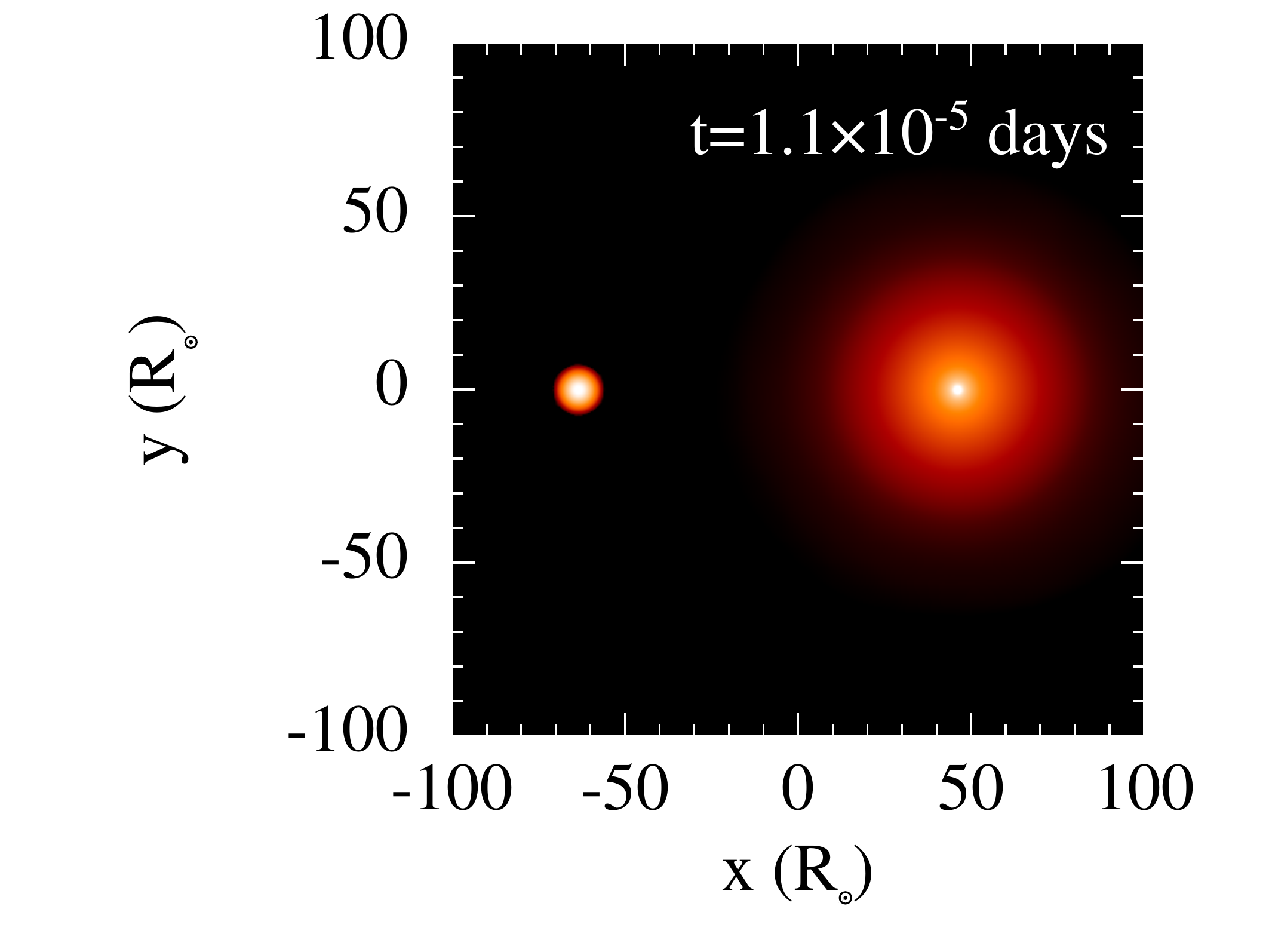}
\includegraphics[scale=0.305, trim={4cm 0cm 1cm 0},clip]{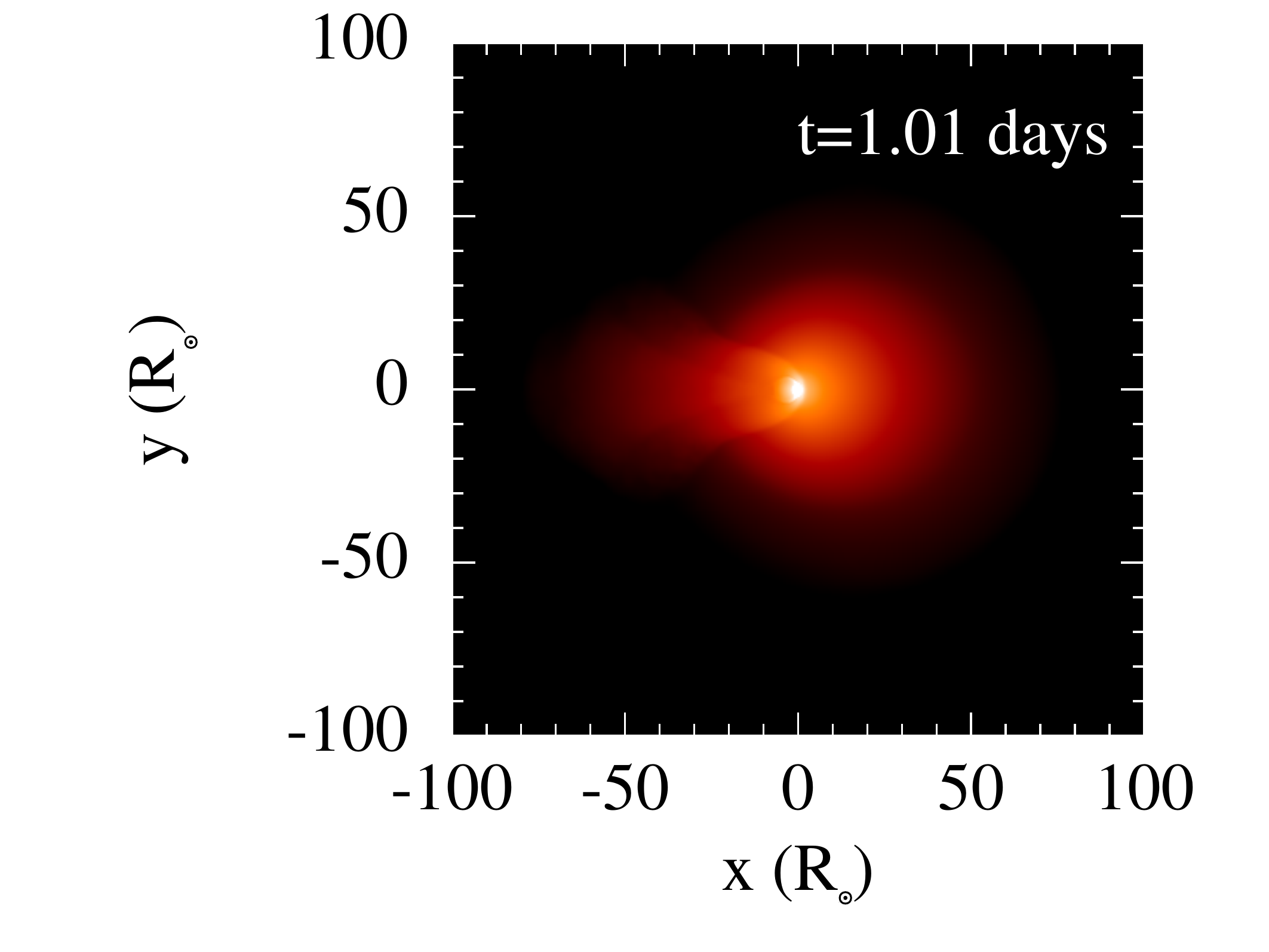}
\includegraphics[scale=0.342, trim={2.5cm 1.05cm 0 0},clip]{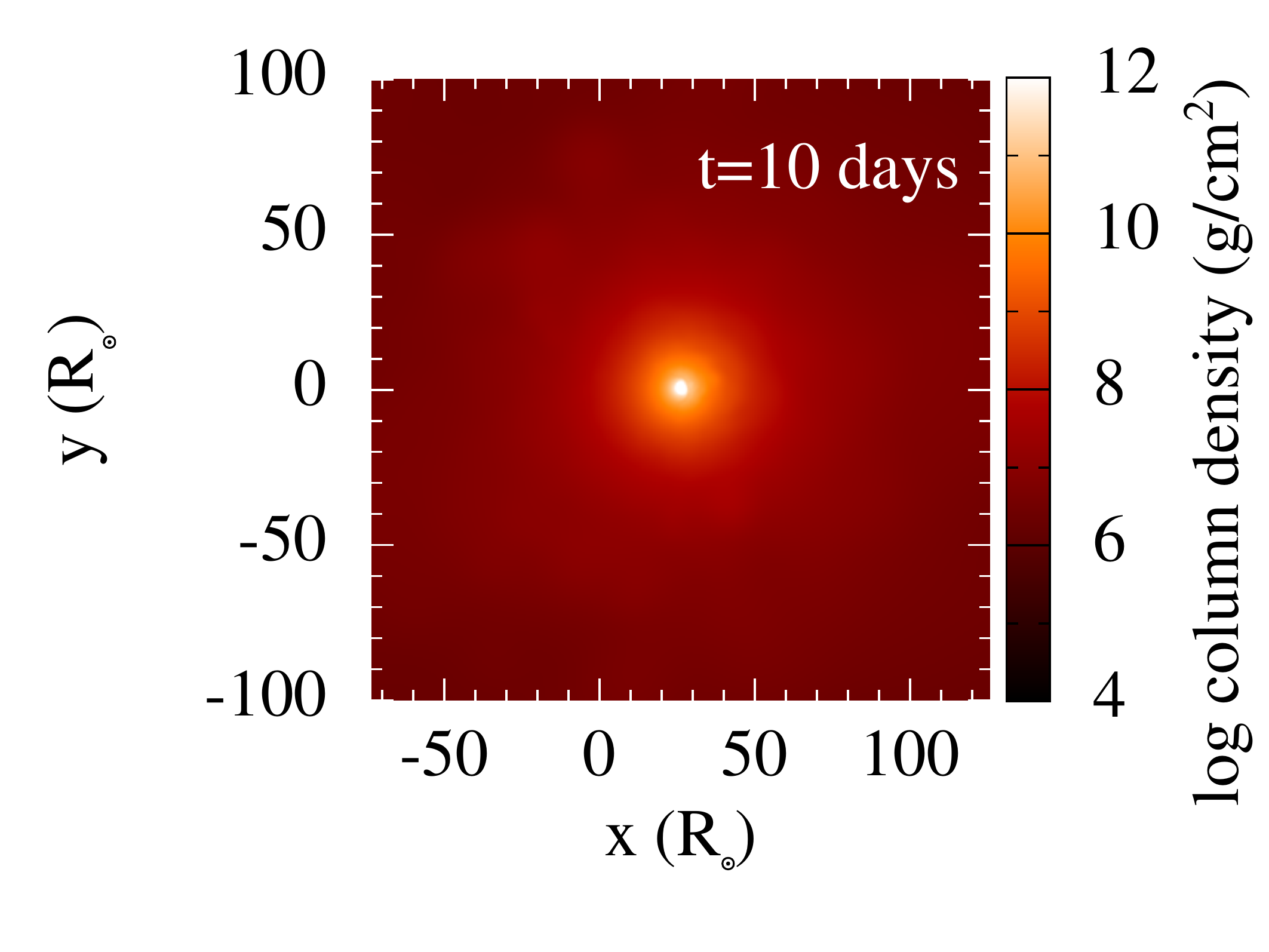}
\caption{Column mass density maps for our hydrodynamical simulation, at the beginning (left-hand panel), at the time of the collision (central panel) and at the end of the simulation (i.e., after 10 days of evolution; right-hand panel). Initially, the two stars are at a distance of $d_{\mathrm{init}}=110$~R$_{\odot}\gtrsim 2\,{} R_{\mathrm{CHeB}}$. As the two stars move on their radial orbit, the secondary enters the much larger envelope of the primary, forming a long cometary tail, and gets disrupted when reaching the inner parts of the CHeB star, after about 1 day of evolution. At the end of the simulation, the stellar remnant relaxes to a much larger envelope, generated by the inflation of the outer layers of the primary star.}\label{denrend}
\end{figure*}

\begin{figure}
\includegraphics[scale=0.4, trim={0cm 1.05cm 0 0},clip]{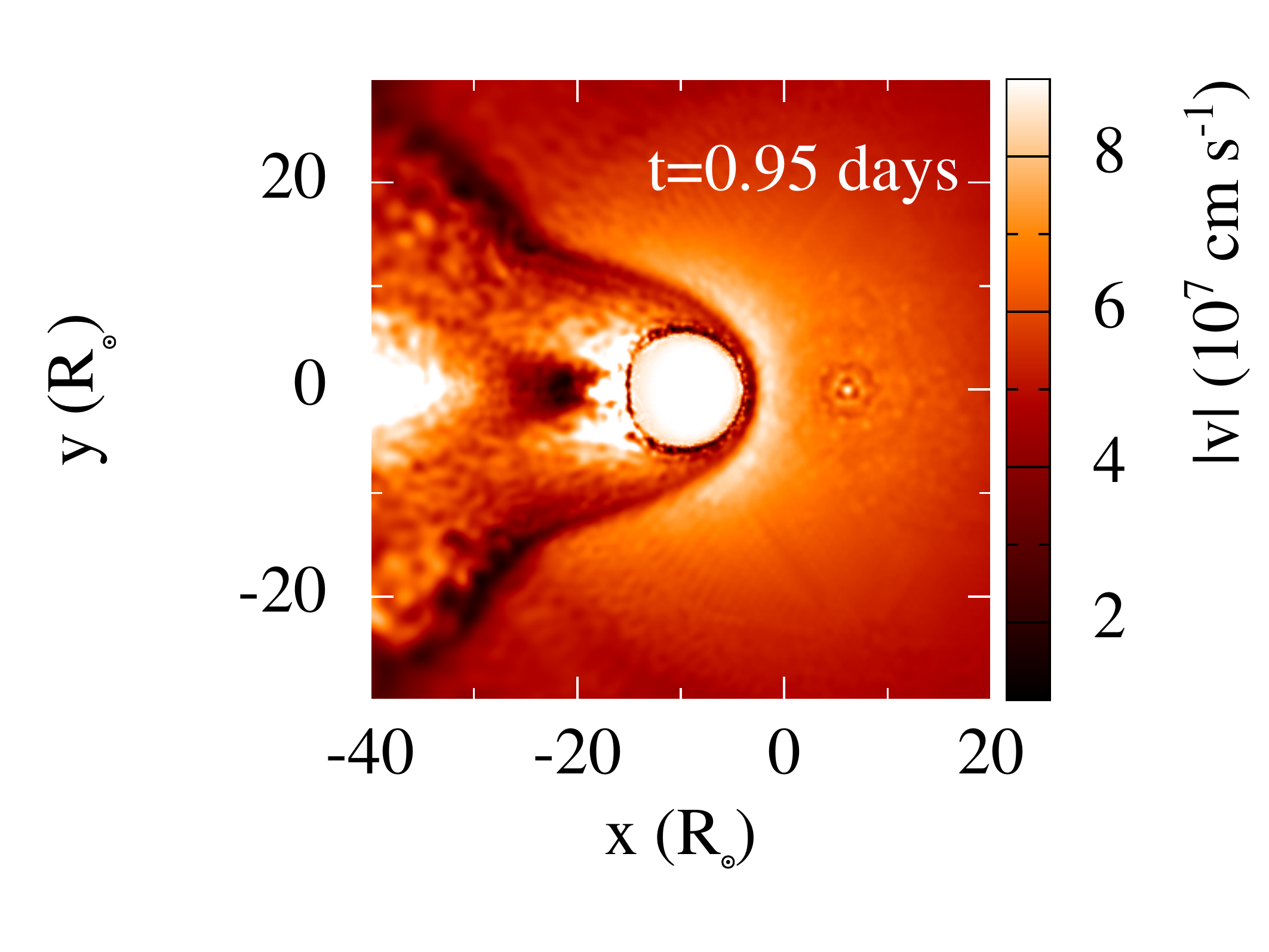}
\caption{Velocity map in a thin slice enclosing the orbital plane at $t = 0.95$ days. A clear leading shock is visible, forming in the envelope of the primary star, as well as a turbulent cometary wake. The white round feature in the center is the bulk of the MS star.}\label{velmap}
\end{figure}

\section{Methods}\label{methods_sec}

In order to compare our results with the ones of \cite{DiCarlo20} and \cite{Renzo20b}, we simulate a specific system, found in the simulations by \cite{DiCarlo20}, which matches the primary black hole mass of GW190521. In the dynamical simulations of \cite{DiCarlo20} (see their Figure 7), a black hole of 88 M$_{\odot}$ is produced by the remnant of the collision between two massive stars, namely:
\begin{itemize}
    \item a core helium burning (CHeB) star, with total stellar mass $M_{\mathrm{CHeB}} = 57.6$~M$_{\odot}$ and radius $R_{\mathrm{CHeB}} = 52.5$~R$_{\odot}$;
    \item a main sequence (MS) star, with total stellar mass $M_{\mathrm{MS}} = 41.9$~M$_{\odot}$  and radius $R_{\mathrm{MS}} = 7.3$~R$_{\odot}$.
\end{itemize}

This specific system is interesting for several reasons. First, the total mass of the two stars is about 100 M$_{\odot}$, so right within the pair-instability mass gap. Furthermore, the two stars are both massive, hence the secondary star could have a strong impact on the collision, compared to a low-mass star. At the same time, the  evolved primary star has a very large radius, and a lower envelope binding energy, compared to a MS star. Hence, it is important to study such a combination of masses and evolutionary phases by means of hydrodynamical simulations, since the outer layers of the primary star could be more easily unbound by the collision.

In the original simulation by \cite{DiCarlo20}, these stars are members of a binary system, lying inside a massive young star cluster. In dense star clusters, stars are subject to strong dynamical interactions, especially in the first Myrs of evolution of the cluster. In the simulation by \citet{DiCarlo20}, the orbit of the binary was strongly perturbed by the gravitational interaction with a third massive stellar object, which brought the two binary components into a nearly-radial orbit and led to a head-on collision between them (see Appendix \hyperref[app_nbody]{A} for more details on stellar collisions in young star clusters).

Previous studies adopting population synthesis  \citep[][]{DiCarlo20} or stellar evolution codes \citep[][]{Renzo20b} cannot estimate the amount of mass lost during the collision and the final chemical mixing of the merger product. In this work, for the first time, we aim at evaluating these two key quantities. 
This is done by means of simulations with the smoothed-particle hydrodynamics (SPH) code {\sc StarSmasher}\footnote{\href{https://github.com/jalombar/starsmasher/}{https://github.com/jalombar/starsmasher/}.} \citep{Gaburov10}, which is particularly suited to model stellar collisions. In its last version, the code implements variational equations of motion and libraries to calculate the gravitational forces between particles using direct summation on NVIDIA GPUs \citep{Gaburov10b}. Using a direct summation instead of a tree-based algorithm for gravity increases the accuracy of the gravity calculations, particularly in terms of energy and angular momentum conservation. At the end of our simulation, the total energy was conserved within a factor $1.5\times 10^{-5}$. 
We adopted a cubic spline for the smoothing SPH kernel, as defined in \citet{Monaghan85}, and the smoothing length of each particle adapts during the simulation to always ensure that at least 40 neighbours are enclosed by it. Shocks are treated by means of an artificial viscosity term, coupled to a Balsara prescription \citep[][]{Balsara95} to avoid spurious inter-particle penetration \citep[][]{Gaburov10}. The simulation time-step is calculated on the fly, according to the usual Courant-Friedrichs-Lewy stability conditions \citep[see, again,][for more details]{Gaburov10}.
The thermo-dynamical treatment of StarSmasher is simplified, compared to the one of sophisticated stellar evolution codes, and is based on adopting an equation of state that includes both the contribution of ideal gas and radiation pressure \citep[][]{Lombardi06,Gaburov10}.

To generate initial conditions, {\sc StarSmasher} comes with an internal module to initialize stars from tables containing one-dimensional profiles of their main properties. For our simulations, we obtained the stellar profiles of our two stars using the stellar evolution code {\sc parsec} \citep{Bressan12, Costa19}. Details about the {\sc parsec} setup used for these simulations can be found in a companion paper \citep[][]{Costa22}.  Figure \ref{ics} shows the density (upper panel), H and He abundance (middle panel) and cumulative mass (lower panel) profiles of the two stars. Both stars have metallicity $Z=0.0002\sim{0.01}\,{}{\rm Z}_\odot$.\footnote{Previous studies have already shown that the stellar collision scenario is effective in producing black holes in the pair-instability gap only at very low metallicities, where stellar winds have little impact on the mass-loss of the stellar remnant, before it collapses into a massive black hole \citep{DiCarlo20b}.} As visible in Figure \ref{ics}, we chose a CHeB primary which has a well developed core, in which part of the helium has already been converted into carbon and oxygen. 
The structure of our primary star is 
the same as in \cite{DiCarlo20b}, while it is different from the one considered by \citet{Renzo20b}, who assumed a primary star at the end of its MS, i.e. with a negligible fraction of CO in the core. 
{\sc StarSmasher} re-samples these one-dimensional profiles with SPH particles distributed in the three-dimensional space on a hexagonal close-packed lattice (to ensure numerical stability), by keeping the number density of particles uniform. This feature is particularly important, since the mass density of stars spans several orders of magnitude from their center to their atmospheres (Fig. \ref{ics}). In this way, particles with different masses can spatially sample each layer of the star in a spatially uniform way. 
In other words, the user decides the number of SPH particles, while {\sc StarSmasher} decides the mass of each single particle. We chose to sample the CHeB and MS star with $8\times 10^5$ and $9\times 10^4$ particles, respectively. Such big difference in resolution is forced by the much wider dynamical range of mass density spanned by the CHeB star and by its much larger radius. As a result, the mass resolution of SPH particles ranges from $\approx 2\times 10^{-11}$ to $\approx 7.6$~M$_{\odot}$ and from $\approx 4\times 10^{-7}$ to $1.4\times10^{-2}$~M$_{\odot}$ for the CHeB and MS star, respectively. 

When importing the stellar profiles, {\sc StarSmasher} checks whether the central density is four times higher than the average density of the stars; in that case, it sets a central core particle.
For this reason and as a result of {\sc StarSmasher}'s algorithm for the three-dimensional distribution of particles, the core of our CHeB star is sampled by a central core particle, that interacts with the other SPH particles only gravitationally, with mass equal to 7.6 M$_\odot$ and smoothing length equal to 1 R$_{\odot}$ and six SPH particles surrounding it (at about 0.35 R$_\odot$) with mass of about 3.5~M$_\odot$ and smoothing length equal to 0.9~R$_\odot$. Setting these large smoothing lengths (of the order of the core size) avoids spurious numerical effects due to gravitational perturbation of the surroundings by these central massive particles. Furthermore, while not being well sampled spatially, most of the mass of the core is sampled by these six massive SPH particles, whose thermal pressure prevents outer particles from spuriously penetrating the core.

Before initializing the collision simulation, the two stars were evolved singularly in relaxation runs, to allow the SPH particles to re-adjust to equilibrium \citep[for more details, see][]{Gaburov10}.

After these preliminary steps, we put the two stars on a hyperbolic radial orbit, with velocity at infinity $v_{\infty}= 10 \;\mathrm{km \;s^{-1}}$ and initial separation of $d_{\mathrm{init}}=110$~R$_{\odot}\gtrsim 2\,{} R_{\mathrm{CHeB}}$, where $R_{\mathrm{CHeB}}$ = 52.5 $R_\odot$ is the radius of the CHeB star. Our choice of $v_{\infty}$ matches the velocity dispersion in young massive star clusters, such as those simulated by \citet{DiCarlo20}. Nonetheless, the value of $v_{\infty}$ adopted here is much lower than $\sqrt{2\,{}G\,{}(M_{\mathrm{CHeB}}+M_{\mathrm{MS}})/d_{\mathrm{init}}}\approx 600\; \mathrm{km\; s^{-1}}$, so it has very little impact on the collision. We assumed an impact parameter equal to zero because we wanted to probe the most extreme case in terms of kinetic energy of the collision and to obtain an upper limit to the mass loss. In follow-up studies, we will explore different orbital configurations.

 While the total permeation of the two stars occurs in about 1 day of simulated time, we decided to evolve our model for 10 days. This timescale is sufficient to reach hydro-static equilibrium, on a dynamical, or sound-crossing, timescale. The thermal relaxation of the new star occurs over a much longer time, that is several times the Kelvin-Helmholtz timescale of the remnant. Three-dimensional hydrodynamical simulations, such as the one presented in this study, are too computationally expensive and lack a proper treatment of several physical processes (e.g. detailed energy injection, transfer and loss) that 
 govern 
 stellar evolution across the whole stellar lifetime. 
 So, the further evolution of the collision product can be reliably achieved only by one-dimensional stellar evolution codes. Hence, in a companion paper \citep[][]{Costa22}, we use the results of the present study to inform stellar evolution simulations of the evolution of the collision product until the end of its life.

\begin{figure*}
\includegraphics[scale=0.6]{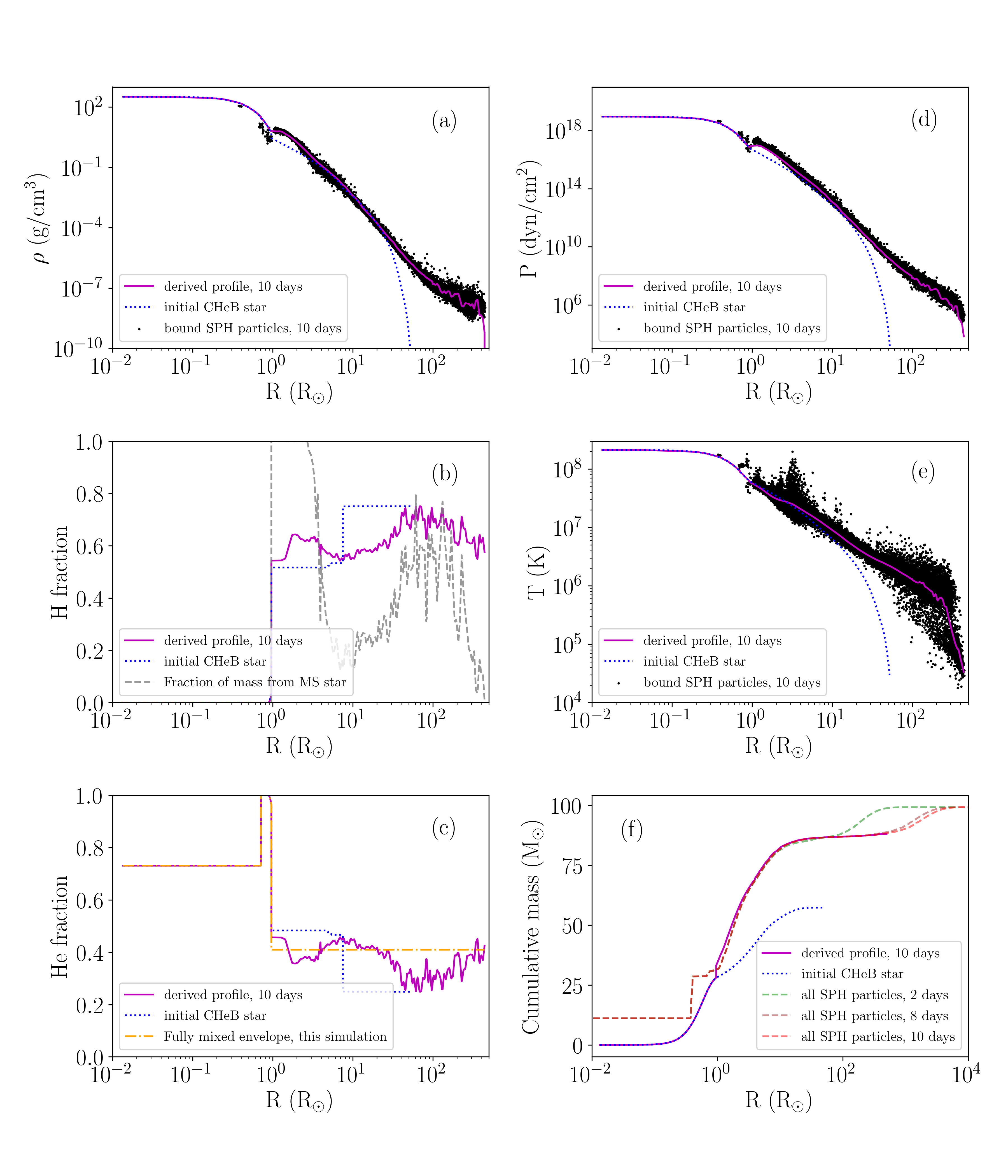}
\caption{Density (panel a), hydrogen (b) and helium (c) mass fractions, pressure (d), temperature (e) and cumulative mass (f) profiles for the post-collision star at the end of the simulation (10 days; solid magenta lines). 
We consider only SPH particles that are still bound to the remnant (black dots, see eq.~\ref{bound} for details). As a comparison, we also show the original profiles for the CHeB star (blue dotted lines). In the middle left-hand panel we also plot the fraction of mass, in each bin, coming from the original MS star (gray dashed line). In the lower left-hand panel we also show the He abundance for a fully mixed envelope (orange dot-dashed line, see text). Finally, in the lower-right panel, we also compare the remnant profile (magenta) to the profile of all the SPH particles (bound+unbound) at a simulation time of 2 (green dashed line), 8 (brown dashed line) and 10 days (red dashed line)}.\label{derprof}
\end{figure*}

\begin{figure*}
\includegraphics[scale=0.55, trim={1cm 0.3cm 0cm 0cm},clip]{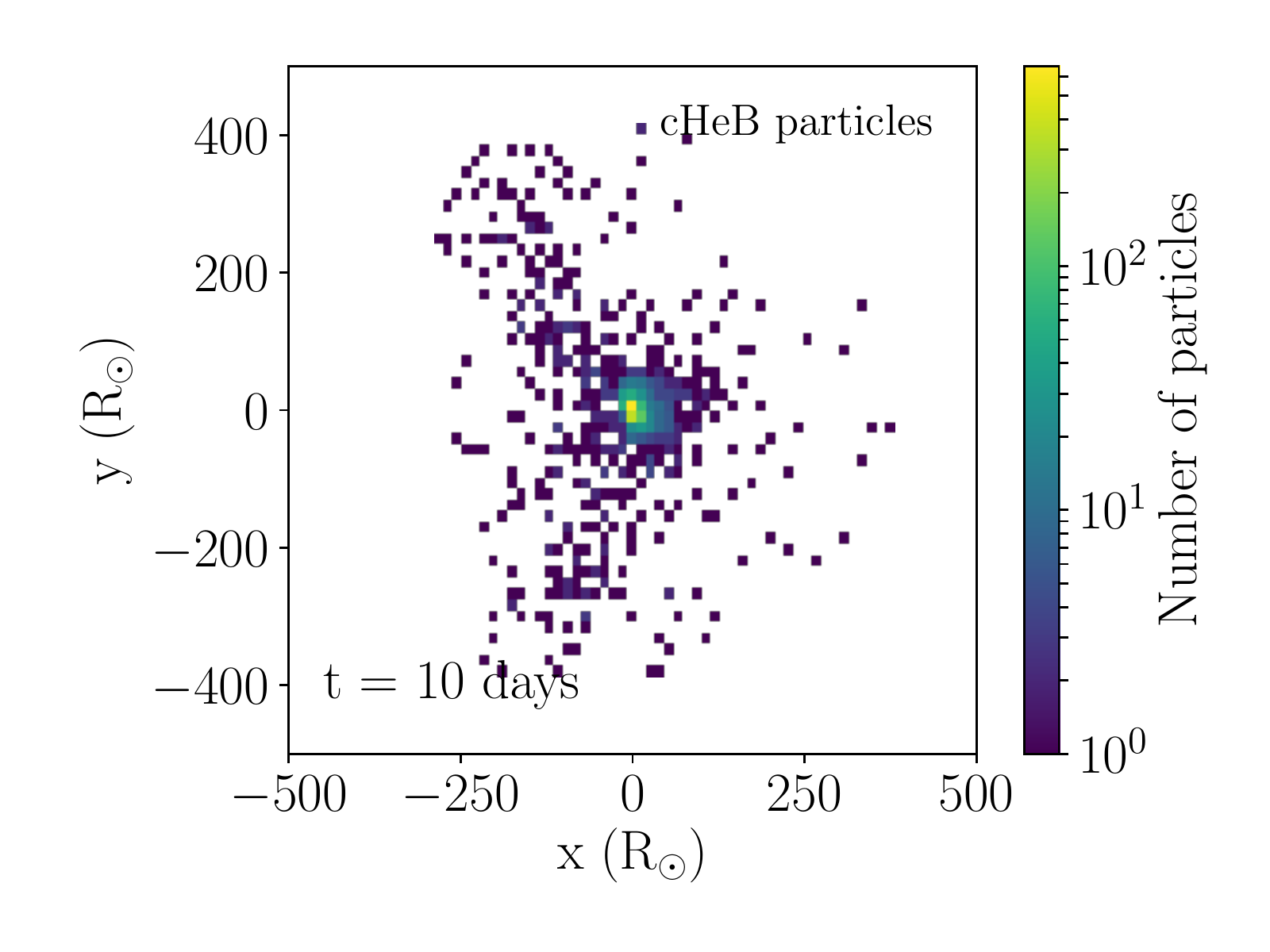}
\includegraphics[scale=0.55, trim={1cm 0.4cm 0cm 0},clip]{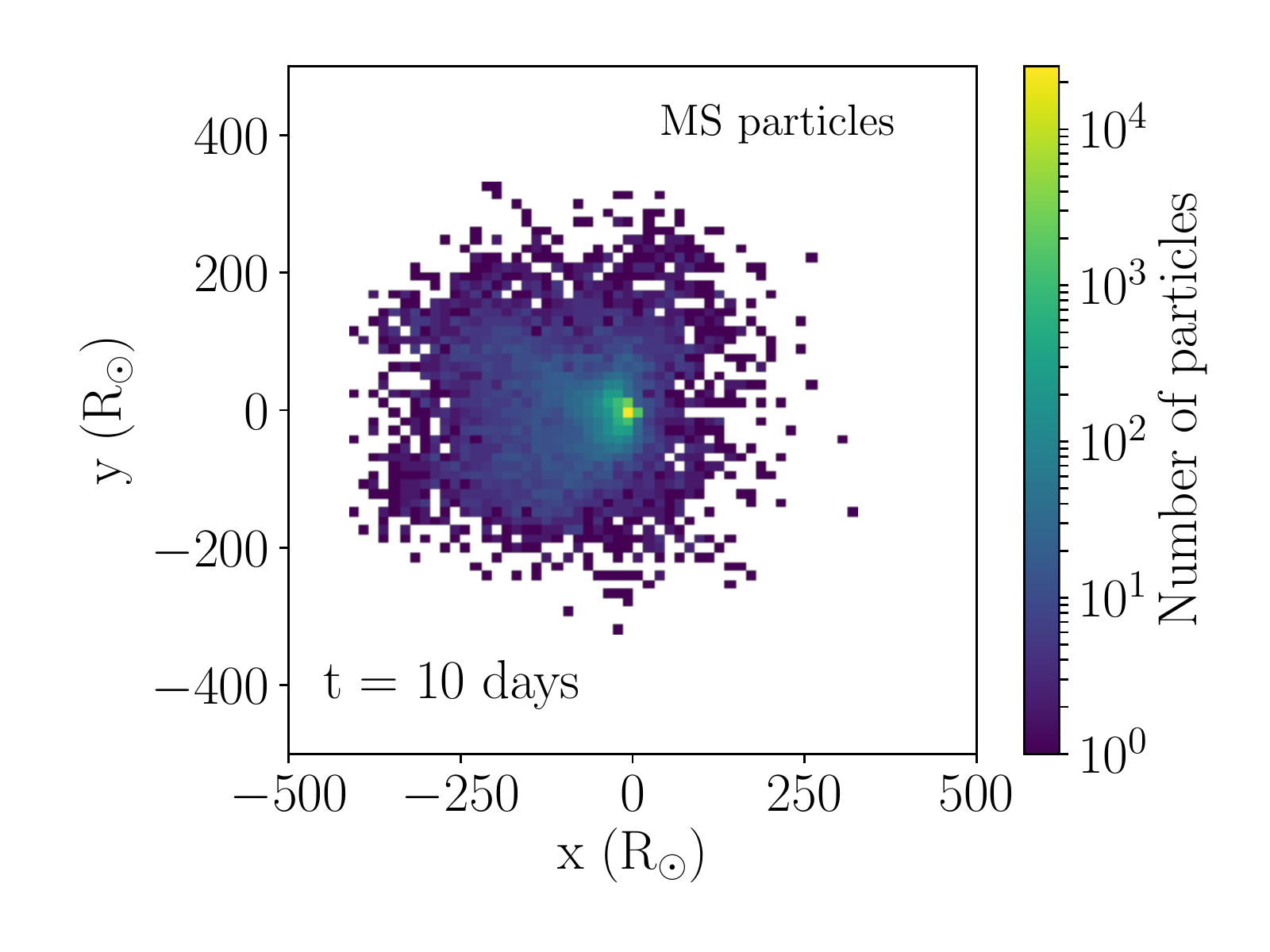}
\caption{Two-dimensional histogram, at the end of the simulation, of the x--y distribution of SPH particles originally belonging to the CHeB (left) and MS (right) stars. Because of our choice of a perfectly head-on collision, both particle subsets are asymmetrically distributed, mostly occupying the side from which their original parent stars reached the collision point.}\label{partmix}
\end{figure*}

\begin{figure*}
\includegraphics[scale=0.55, trim={1cm 0.4cm 0cm 0},clip]{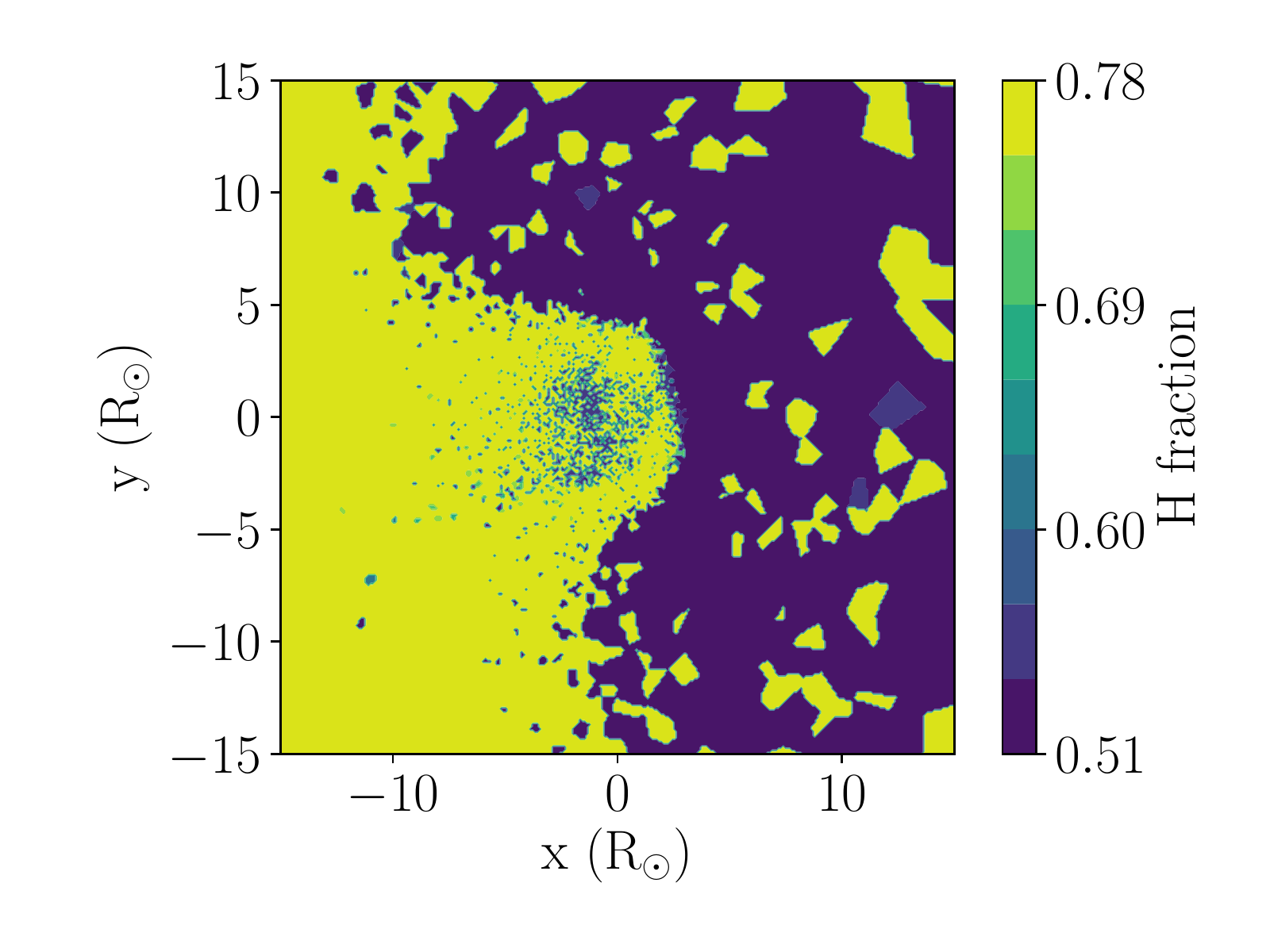}
\includegraphics[scale=0.55, trim={1cm 0.4cm 0cm 0},clip]{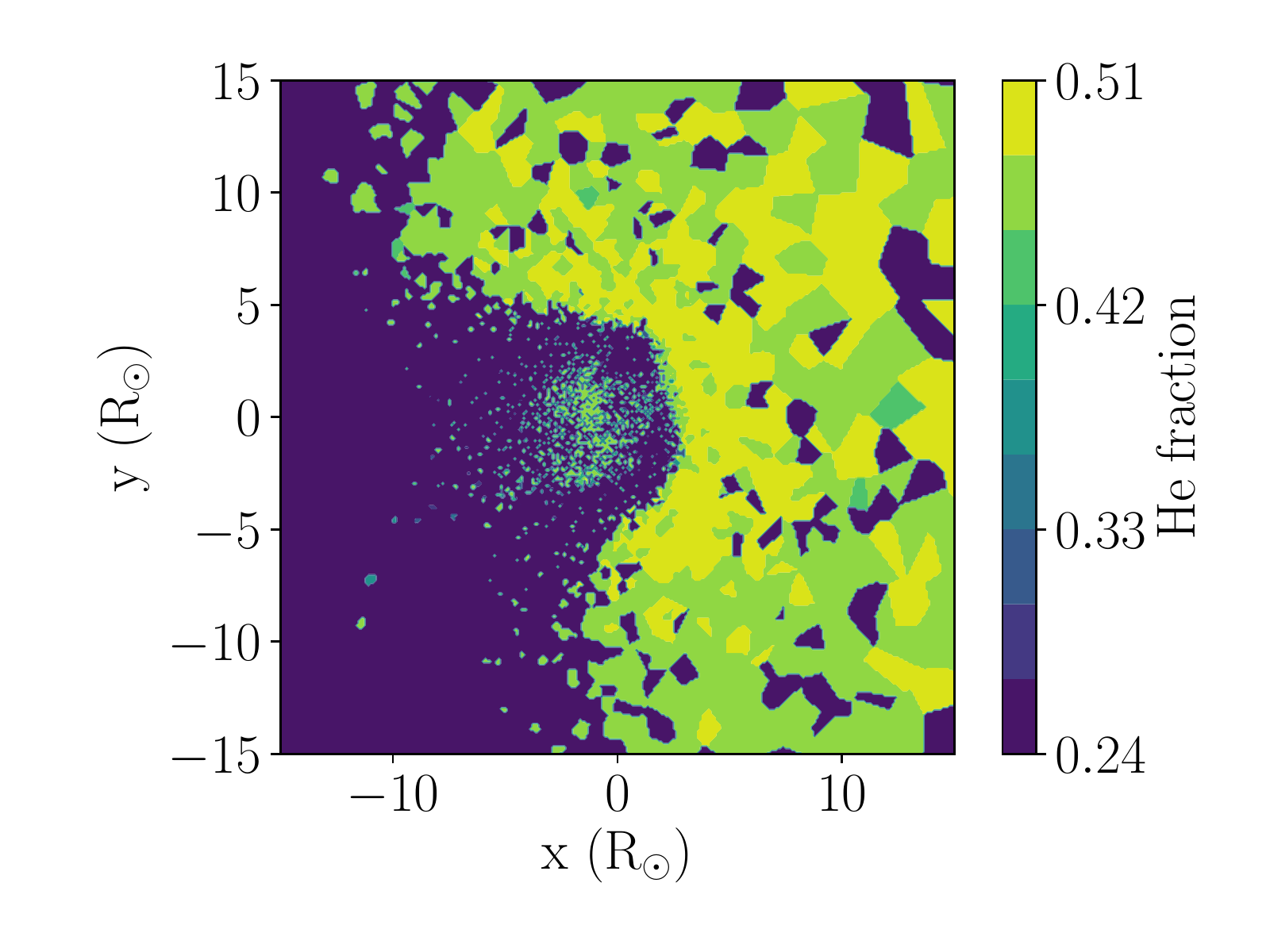}
\caption{Voronoi tessellation maps of the hydrogen (left) and helium (right) mass fractions, projected in the x--y domain. The particle distribution asymmetry is also reflected in the final distribution of chemical abundances.}\label{partabund}
\end{figure*}
\section{Results}

Figure~\ref{denrend} shows the evolution of the collision at the beginning of the simulation, during the maximum permeation of the two stars and at the end of the simulation. As the MS star plunges in the atmosphere of the CHeB star, its outer layers form a strong shock in the frontal side of the collision, while they lead to a cometary tail in the back side  (Figure \ref{velmap}). When the MS star reaches a separation, relative to the core of the CHeB star, of the order of its Roche limit (i.e., about the star's initial radius, in the simple Roche approximation), it is tidally disrupted by the core of the CHeB star. After the coalescence, the remnant star shows a much more extended envelope, generated by the inflation of the outer layers of the primary star due to conversion of the kinetic energy of the impact into thermal and kinetic energy of those layers.

Figure~\ref{derprof} shows density, H and He abundances, pressure, temperature and cumulative mass profiles for the stellar remnant, obtained through a mass-weighted average of these quantities over spherical shells centered on the core of the final star. 
In order to calculate these profiles, we 
estimated which particles are still bound to the remnant, with a simplified approach. For each $i-$th SPH particle, we calculated \begin{equation}\label{bound}
    e_{\rm i}=v_{\rm cm,i}^2+u_{\rm i}-GM_{\rm encl,i}/d_{\rm cm,i},
\end{equation}
where $e_{\rm i}$ and $u_{\rm i}$ are the specific total and internal energy, $v_{\rm cm,i}$ and $d_{\rm cm,i}$ are the velocity and distance calculated with respect to the total center of mass, and $M_{\rm encl,i}$ is the total mass enclosed within $d_{\rm cm,i}$. Every particle with positive $e_{\rm i}$ is considered unbound and  is excluded from the calculation of the remnant profiles.  
We assumed that the core of the remnant has exactly the same properties as the core of the original CHeB star, hence the profiles in the inner 1 $R_{\odot}$ are just taken from the initial {\sc parsec} profiles of the primary, used to initialize our simulation.

As visible particularly in the cumulative mass profiles (lower right-hand panel of Fig. \ref{derprof}), the SPH particles re-adjust to a new mass distribution that is more concentrated compared to the one of the initial CHeB star, but that extends up to a much larger radius of $\approx 450\,{}$~R$_{\odot}$. The lower right-hand panel also shows that the unbound mass is in a large bulk of particles that are launched at much larger radii ($>1000$~R$_{\odot}$) already 8 days after the collision. 

From this calculation, we got that the mass of the collision product is $\approx 87.9$~M$_{\odot}$, i.e. about 11.7\% of the initial stellar mass is lost in the impact. In particular, the final star  retains about 82\% and 98\% of the mass of the CHeB star and MS star, respectively. Hence, the structure of the CHeB expands in the collision, with its shock-heated outer envelope ending up unbound.

In order to calculate abundances profiles, we assigned each particle at the beginning of the simulation the hydrogen and helium mass fractions of the shell of the original stars  they initially belonged to. In this way, chemical abundances are simply advected by the SPH particles, in a Lagrangian fashion. 
The post-collision abundance profiles show that:
\begin{itemize}
    \item  the MS star disrupts outside the denser core of the CHeB star, depositing most of its material in a shell surrounding the core  (1~R$_{\odot} \lesssim d_{cm} \lesssim  5$~R$_{\odot}$), enriching this region with hydrogen;
    \item some ram-pressure stripped and shocked outer layers of the MS star are deposited further out in the envelope. As a result, at larger distances from the center ($d_{cm} \gtrsim  5$~R$_{\odot}$), the profile shows an increase in  H, with a maximum H-enrichment at about 100~R$_{\odot}$ (similar to the original radius of the  CHeB envelope);
    \item the collision expels part of the envelope of the CHeB star and the outer layers of the remnant are mostly composed of material originally belonging to the H-burning shell of the CHeB star.
\end{itemize}

In particular, the abundance profiles surrounding the core of the remnant  almost perfectly match those of the original MS star (Fig. \ref{ics}). This happens because the shell between 1 and 5 R$_{\odot}$ is totally composed of material originally belonging to the MS star (gray dashed line in the central left-hand panel of Fig. \ref{derprof}). With a similar argument, the He at large radii is material belonging to the He-shell of the original CHeB star that is brought at larger radii by the impact. The MS star pollutes the outer envelope mostly between 40 and 200 R$_{\odot}$, leading to the dip in He abundance in that region and leaving higher He fractions in the outer $\approx 300$ R$_{\odot}$ of the newly formed star. 

For comparison, were we assuming a complete mixing of the MS with the CHeB original envelope, we would have obtained a constant He fraction in the envelope of the new star equal to 0.41 (orange dash-dotted line in the lower left panel of Fig.\ref{derprof}). This is a lower He fraction than assumed by \citet{Renzo20b}, due to the different structure of the two initial stars (see Sec. \ref{methods_sec}). A somewhat exotic He fraction in the envelope would imply a different position of the stellar remnant in the Hertzsprung-Russell diagram and peculiar line absorption features of its spectrum.

By looking at the three-dimensional distributions, our simulation shows that significant asymmetries arise in the spatial profile and chemical composition. Especially at large radii, particles originally belonging to the MS star are distributed along the radial stellar orbit, with the shape of a cometary tail (Fig. \ref{partmix}). Figure \ref{partabund} also clearly shows that the fresh hydrogen brought in the envelope of the post-coalescence star is mostly one-sided, except for the closest vicinity to the original CHeB core. We report spherically-averaged quantities in Fig. \ref{derprof}, because we expect any rotation imparted by a non-zero impact parameter to azimuthally redistribute this material over a time-span much faster than the evolutionary timescale of the star.

\section{Discussion}

Our simulation shows that up to 12\% of the CHeB star's envelope mass can be lost in a collision similar to the one described in \citet{DiCarlo20}. Our assumption of a head-on collision must be regarded as an upper limit to the mass loss.  

Our result is in agreement with the relation obtained by \citet{Lombardi02}; according to their findings, the fraction $f$ of mass loss in our head-on collisions should be 
\begin{equation}
    f= c_{\rm 1}\frac{q}{(1+q)^2}\frac{R_{\rm CHeB,0.86}+R_{\rm MS,0.86}}{R_{\rm CHeB,0.5}+R_{\rm MS,0.5}},
\end{equation}
where $c_{\rm 1}=0.157$, $q=M_{\rm MS}/M_{\rm CHeB}$ and $R_{\rm CHeB,0.86}\approx 6.64$~R$_{\odot}$ ($R_{\rm CHeB,0.5}\approx 1.06$~ R$_{\odot}$) and $R_{\rm MS,0.86}\approx 3.47$~R$_{\odot}$ ($R_{\rm MS,0.5}\approx 2.15$~ R$_{\odot}$) are the radii of the CHeB and MS stars containing the 86\% (50\%) of their total mass. When applying this formula to our initial conditions, we found that the predicted mass loss is  12\%.

We can also compare our simulation to an approximation that is frequently used in the literature to derive the structure of stars formed in gentle (i.e., shock-free) mergers \citep[see, e.g.,][]{Lombardi96, Gaburov08,Glebbeek13}, where the post-collision particles are assumed to arrange radially with increasing initial specific entropy. Before and after the collision, we calculate the entropic variable (or buoyancy), $A$, of each particle, defined as \citep{Gaburov08}
\begin{equation}
    A=\frac{P_{\rm tot}}{\rho^{5/3}}\exp{[8\,{}(1-\beta)/\beta]},
\end{equation}
where $P_{\rm tot}=P_{\rm gas}+P_{\rm rad}$ is the total pressure, including both the gas and the black-body radiation pressure, $\rho$ is the gas density and $\beta=P_{\rm gas}/P_{\rm tot}$. The entropic variable $A$ is a function of the fluid specific entropy and chemical composition \citep{Gaburov08}. In the approximation of no heating (e.g., in the absence of shocks) or mixing between them, one expects the different fluid elements of the two stars to rearrange into hydrostatic equilibrium, with $A$ monotonically increasing outward. 

\begin{figure}
\includegraphics[scale=0.55]{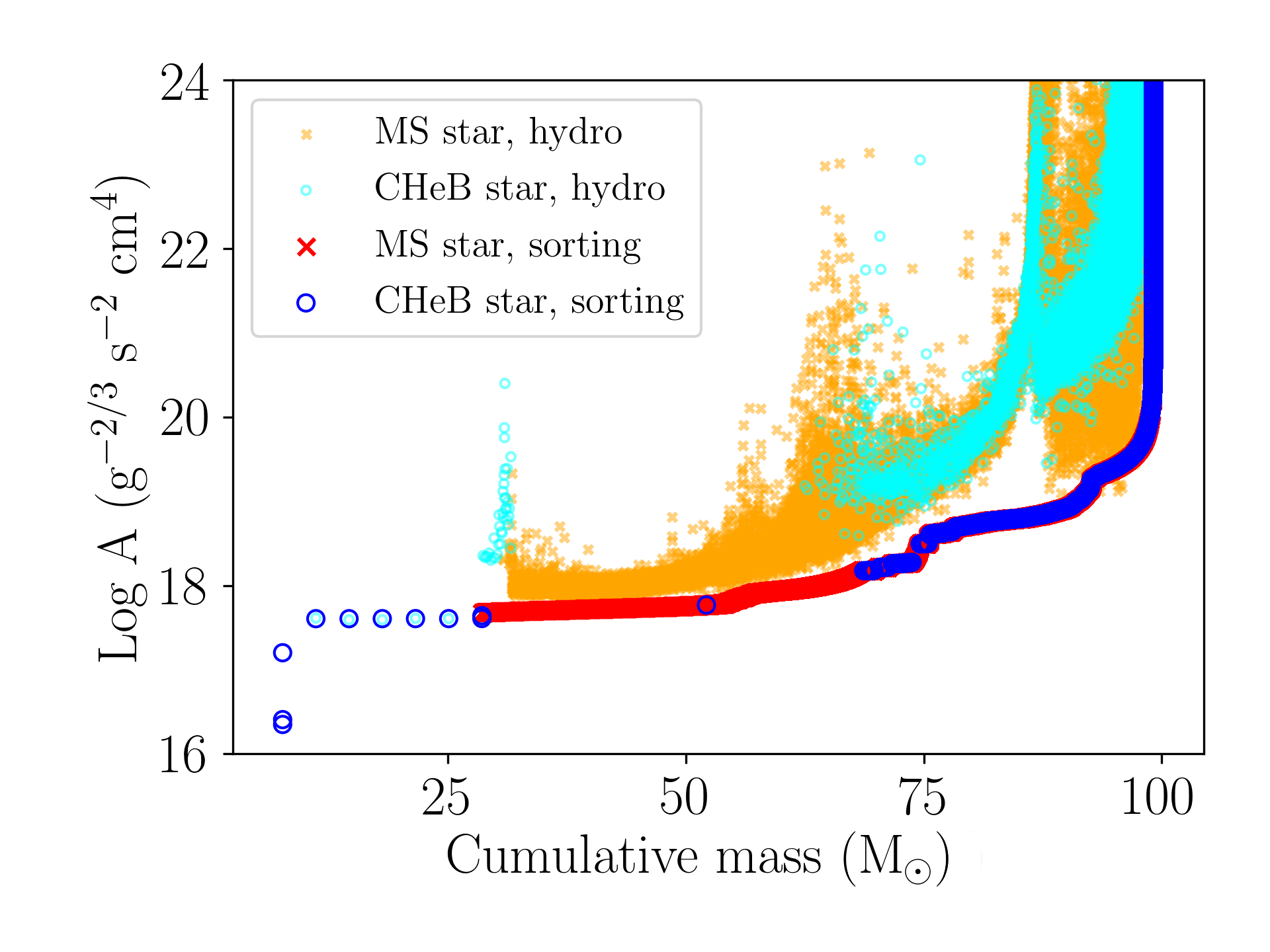}
\caption{Entropic variable $A$ as a function of cumulative mass, for our initial conditions sorted by increasing value of $A$ (large blue circle and large red crosses) for the CHeB and MS star, respectively) and for the output of our hydrodynamical simulation (small cyan circles and small orange crosses for the CHeB and MS star, respectively). For a gentle, shock-free collision, the post-collision SPH particles (small cyan and orange markers) are expected to rearrange according to the sorted particles (large blue and red markers), with the MS sinking to the core of the cHeB star, without penetrating it. The stratification of material in our hydrodynamical simulation shows a  good agreement with this simple prediction, despite the presence of shocks in our simulation.} \label{entropy}
\end{figure}

Figure \ref{entropy} compares the $A$ distribution (in cumulative mass) of our initial SPH particles, when sorted by increasing $A$, to the actual profile of the SPH particles at the end of our simulation. Sorting the particles of the initial colliding stars by increasing $A$ predicts that the MS star would sink down to the outer boundary of the CHeB core even in the case of a less extreme collision, with non-zero impact parameter. This is in 
good agreement with the conditions at the end of our simulation, which are also plotted in Figure \ref{entropy}. A large portion of the MS star sinks to near the core of CHeB star with relatively small changes in $A$, indicating that much of the MS material is not strongly affected by shocks in the plunge.
The entropy sorting also confirms that the core of the primary star is not penetrated by the secondary in the collision, because of its lower specific entropy. As already discussed in the literature \citep[e.g., case P in][]{Glebbeek08, Glebbeek13}, when the primary star is in a sufficiently late stage of its evolution, its core becomes the core of the collision product.

However, traces of shocked material that significantly modify the initial entropy profile are also present in Figure \ref{entropy}. These are especially clearly seen in the specific entropy of the outer $\sim 25$~M$_\odot$, and the distribution of shock-heated and ablated MS material within those outer layers. This shock-heating has the effect of rearranging the post-collision structure relative to what might be predicted by the initial entropy sorting (in one extreme) or completely homogeneous mixing (in the other extreme). Thus, the details of the shock-heating in the collision process are crucial in shaping the final post-coalescence composition shown in Figure \ref{denrend}.

We conclude, independently of our zero-impact parameter orbital configuration, that the envelope of the primary star can be strongly enriched by material of the secondary (in our case, a large amount of fresh hydrogen) in the vicinity of the CHeB core. Physically, this arises because of the density and specific entropy contrast between the MS star and the CHeB envelope. 
In their post-collision model, \citet{Renzo20b} assume that the MS star mixes with the envelope of the more evolved star homogeneously with radius, with a uniform He fraction of 0.52 in the envelope of the collision product\footnote{This specific value depends on the specific choice of the properties of the colliding MS star.}. Our simulation instead predicts a variable H/He abundance depending on the distance from the central core.

The expelled mass largely originates into the outer, H-rich envelope of the CHeB star. As a consequence, the outer envelope of the stellar remnant has higher He abundances compared to those of the outer layers of the original CHeB star, 
reaching values of the order of 0.4 at the surface.

In a companion paper \citep[][]{Costa22}, we carefully model the evolution of a star with same mass and chemical abundance profiles of our model, confirming that it can collapse to a black hole directly, though the final mass of the black hole is sensitive to the fraction of the envelope that can be ejected during a failed supernova \citep[e.g.,][]{fernandez2018}.
 
As a possible caveat, a larger impact parameter could reduce the amount of mass lost in the impact and would induce differential rotation in the collision product \citep[e.g.,][]{rasio95,Sills01,schneider19}. This could lead to higher mixing in the envelope and have some impact on the evolution of the remnant.
The initial velocity at infinity represents another important assumption: here, we chose $v_{\infty}=10$ km s$^{-1}$ to mimic collisions in young stellar clusters. Encounters in more massive star clusters (e.g., nuclear star clusters) involve higher relative velocities ($\approx{50-100}$ km s$^{-1}$). 
Furthermore, we also expect our results to significantly depend on the evolutionary phase of the two colliding stars. We discuss some of these aspects in Appendix \hyperref[app_param]{B}. In a follow-up study, we will systematically focus on these uncertainties, by running a grid of hydrodynamical simulations with a range of impact parameters, relative velocities, and evolutionary phases.

\section{Summary} 

Previous works predict that massive star collisions might lead to the formation of black holes in the pair-instability mass gap \citep[e.g.,][]{DiCarlo19,DiCarlo20b,Renzo20b}. This prediction is subject to two main uncertainties: the amount of mass lost during the collision and the level of chemical mixing in the collision product. Here, for the first time, we are able to estimate the mass loss and the chemical mixing, by means of a three-dimensional hydrodynamical simulation of the collision between a primary massive CHeB star and a secondary massive MS star.

With our work, we showed that even in the case of a head-on collision, our stellar remnant has a mass of $\approx 88$~M$_{\odot}$. The mass lost during the collision is 12\% of the initial total mass of the two stars, and is mostly part of the H-rich outer envelope of the CHeB star.

Our model also shows that the MS star dissolves in the envelope of the CHeB star, therefore rearranging the chemical composition of the primary envelope. 
The collision product shows large He abundances at its surface ($\approx 0.4$ for our model), which are a very distinctive observational feature.
Our results confirm that stellar collisions are a viable mechanism to form black holes in the mass gap ($\sim{60-120}$ M$_\odot$), even if the details depend on the late evolution of the stellar remnant \citep[][]{Costa22} and on the mass fraction that can be ejected during a failed supernova \citep[e.g.,][]{fernandez2018}.

\section*{Acknowledgements}

The authors would like to thank Mathieu Renzo for his very useful and detailed feedback. 
AB, GC and MM acknowledge financial support by the European Research Council for the ERC Consolidator grant DEMOBLACK, under contract no. 770017. We acknowledge the CINECA award HP10CCQZCO under the ISCRA initiative and the CINECA-INFN agreement, for the availability of high performance computing resources and support. M. MacLeod acknowledges support by the US National Science Foundation under Grant No. 1909203.  The initial conditions were obtained with the stellar evolution code {\sc parsec} \citep{Bressan12, Costa19} and the hydrodynamical simulations were run with the SPH code {\sc StarSmasher} \citep[][]{Gaburov10}. The figures and the data analysis were performed either through the free and open source visualisation tool {\sc splash} \citep{Price07} or through {\sc Python}'s libraries {\sc NumPy} and {\sc matplotlib}.

\section*{Data Availability}

Data underlying this article are publicly available at the following link: \href{https://zenodo.org/record/6882301}{10.5281/zenodo.6882301}



\bibliographystyle{mnras}
\bibliography{litmerg} 




\appendix

\section{Stellar collisions in young star clusters}\label{app_nbody}

Understanding how common stellar collisions are in star clusters requires direct N-body simulations with binary population synthesis \citep[e.g.,][]{DiCarlo20b}. Here, we discuss the results of a set of 3555 simulations of star clusters with initial mass between 5000 and 8000 M$_\odot$, and two additional star clusters with initial mass $5\times 10^4 \, \rm{M_{\odot}}$. We ran these clusters with the direct N-body code {\sc nbody6++GPU} \citep{Wang15}, interfaced with the {\sc mobse} population synthesis code \citep{Mapelli17,Giacobbo18}. For more details on the numerical setup of these simulations, we refer to \cite{Torniamenti22}.
\begin{figure}
\includegraphics[scale=0.30]{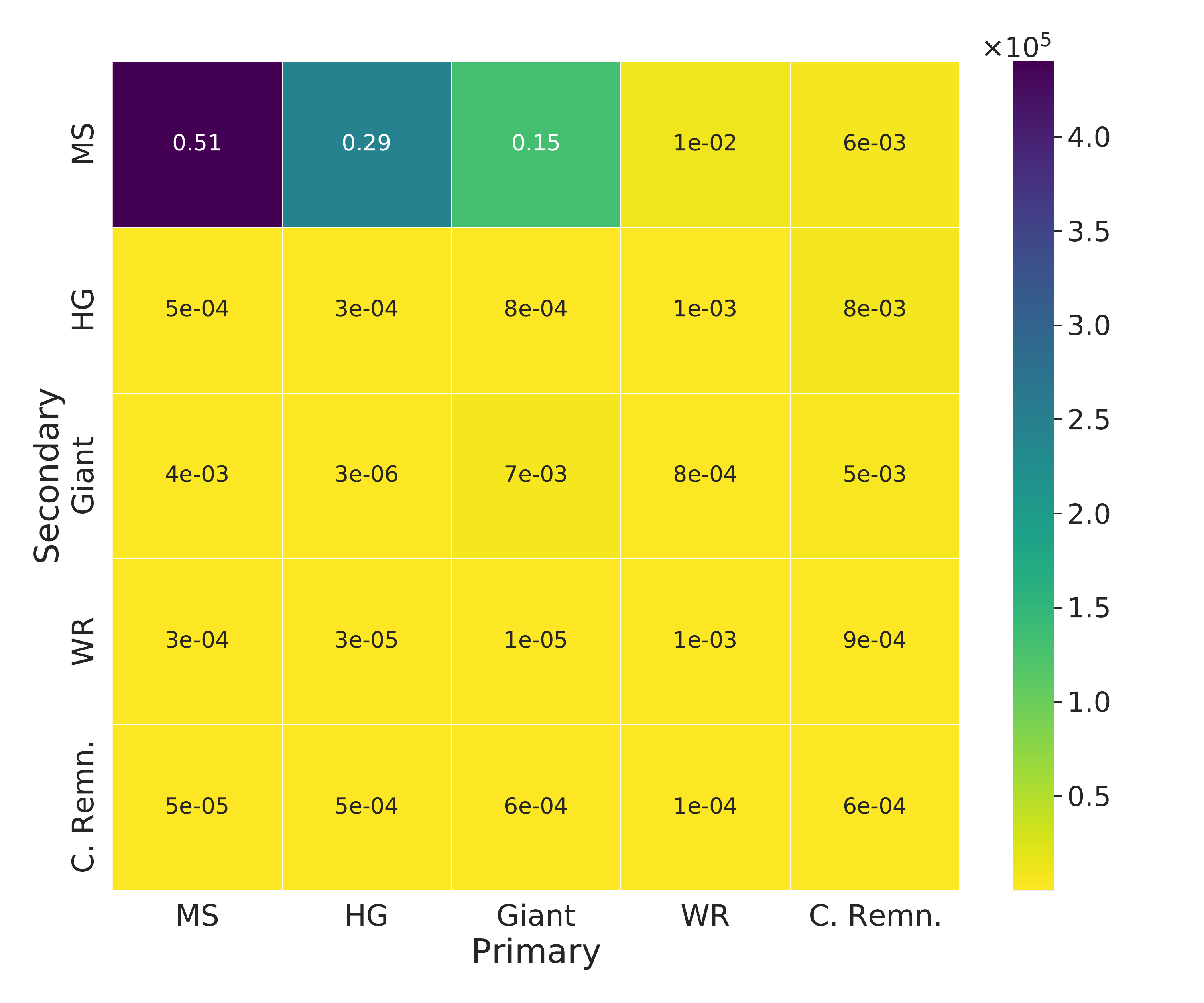}
\caption{Collision matrix from a set of direct N-body simulations of star clusters. The x-axis (y-axis) shows the type of the primary (secondary) colliding stars. The possible types are MS, Hertzsprung gap (HG), CHeB giant (Giant), Wolf-Rayet (WR) star, and compact remnant (C.Remn.). The numbers in the matrix are the fractions of collisions corresponding to a given combination of primary--secondary star type. We find that 51\%, 29\%, and 15\% of all collisions involve a MS, HG, and Giant primary, respectively. The vast majority (96\%) of collisions involve a secondary star that is still on the MS.  The linear colour-map indicates the number of collisions corresponding to a given type in our simulations.}\label{nbody_coll}
\end{figure}

Figure~\ref{nbody_coll} summarizes the key result. The vast majority of star-star collisions involve a MS secondary star (96\% of the cases). The most-common primary member of a star-star collision is a MS (51\% of the collisions). The primary star is a Hertzsprung gap or a CHeB giant star in 29\% and 15\% of the collisions, respectively. The other possible combinations -- i.e., collisions involving Wolf-Rayet stars or compact remnants -- are at least one order of magnitude less common ($\leq{}1$\%). The main case considered in our work, i.e. the collision between a CHeB primary star and a MS secondary star is thus the third most common case. 

The colour map of Figure~\ref{nbody_coll} indicates that we found $\sim{4\times{}10^5},$ $2.5\times{}10^5$ and $10^5$ collisions between two MS stars, one Hertzsprung gap and a MS star, and one CHeB giant and a MS star in our simulations, respectively. They correspond to 1.9, 1.1, and 0.6 collisions every $10^2$ M$_\odot$ of initial stellar mass.

All the collisions found with our direct N-body simulations happen between two members of a former binary system and are triggered by a close dynamical encounter of the binary with an intruder (another star or compact remnant). Such interaction completely perturbs the orbit of the binary, leading to a prompt collision.

\section{Parameter exploration}\label{app_param}

\begin{figure}
\includegraphics[scale=0.58]{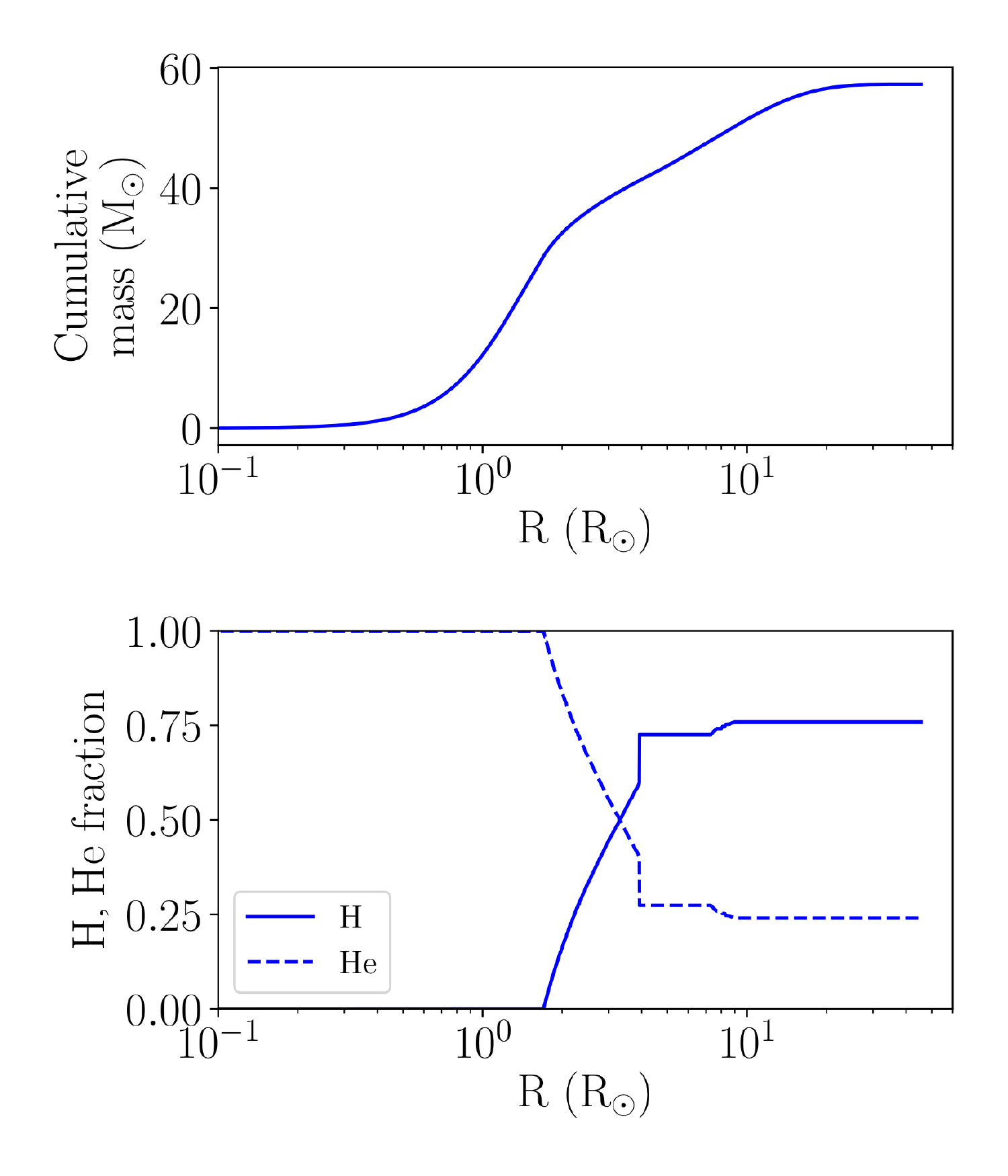}
\caption{Initial cumulative mass (upper panel) and H and He fraction (lower panel) profiles for the primary star of model {\bf primTAMS}.}\label{prof_ics_renzo}
\end{figure}

In this Appendix, we present several additional hydrodynamical simulations with {\sc StarSmasher}. These were obtained by keeping  the same setup of our fiducial model, except for one simulation parameter. The simulations are:

\begin{itemize}
    \item model \textbf{v100} has a velocity at infinity $v_{\rm inf}=100$ km/s. This model probes the effect of a very massive star cluster, with a much higher velocity dispersion (e.g., a nuclear star cluster with velocity dispersion $\approx{100}$ km/s). A third body perturbing the initial orbit of our primary and secondary stars can lead to a more energetic head-on collision.
    \item model \textbf{b0.1}, in which the two stars are on a hyperbolic orbit with pericentre distance $d_{\rm peri}=0.1\; \mathrm{R}_{\odot}$.
    \item model \textbf{primTAMS}, where the primary star is in a younger evolutionary phase, i.e. at its terminal-age MS. For this model, our primary profile is the same {\sc mesa} profile as the primary star in the study by \cite{Renzo20b}, with a core with 100\% abundance of He (Fig. \ref{prof_ics_renzo}). The secondary star is the same adopted for all the other models.
    \item model \textbf{res450}, in which the primary has been sampled with $4.5\times{}10^5$ SPH particles, i.e. slightly more than half of the resolution of the primary in our fiducial model. This model is meant to study the impact of numerical resolution on the main results.
\end{itemize}

\begin{table}
  \centering
    \begin{tabular}{ccc}
    \hline
           model & Final bound mass (M$_\odot$) & Final r$_{\rm max}$ (R$_\odot$)  \\
          \hline\hline
   \bf fiducial & 87.9 & 441.1 \\
   \bf v100 & 87.7 & 434.5 \\
   \bf b0.1 & 87.1 & 434.2 \\
   \bf primTAMS & 90.0 & 454.5 \\
   \bf res450 & 87.2 & 437.5 \\
    \hline
    \end{tabular}  
  \caption{Final bound mass and r$_{\rm max}$ (defined as the position of the farthest bound particle) for all the models presented in this work.}\label{tab_mass_size}
\end{table}

\begin{figure}
\includegraphics[scale=0.55]{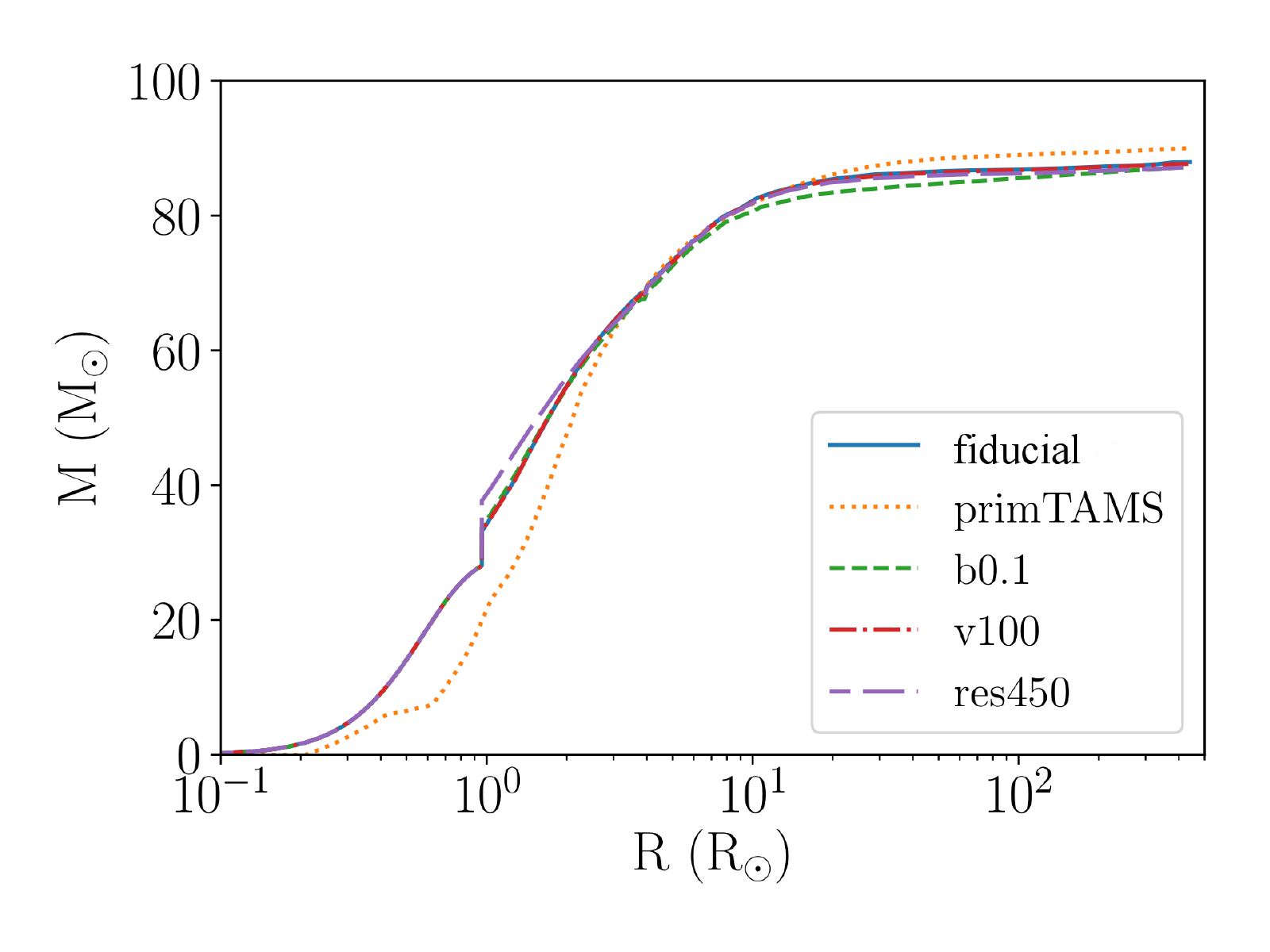}
\caption{Final cumulative mass profiles of our standard model, compared to the models described in this Appendix.}\label{mass_set}
\end{figure}

Figure \ref{mass_set} shows the cumulative mass profiles of the aforementioned models, along with the one of our fiducial model. As visible also in Table \ref{tab_mass_size}, all the models have a comparable final mass of the collision product.

The {\bf res450} model shows that even adopting half the SPH resolution for the primary star gives a final mass that is equal to our fiducial model within $\approx 1 $ \%, thus proving that numerical convergence has been reached.

The collision product mass is consistent also in models {\bf b0.1} and {\bf v100}. For model {\bf b0.1}, the adopted $d_{\rm peri}=0.1 \; \mathrm{R}_{\odot}$ is small enough to give a very similar dynamics of the impact, though the collision product acquires angular momentum from the initial orbit, with its inner 50 $\mathrm{R}_{\odot}$ reaching rotational velocities of the order of 200 km/s \citep[though rotational instabilities can transport angular momentum towards the surface of the remnant during its further evolution, as discussed in][]{Sills01}.

In the case of model {\bf v100}, even with a factor 10 larger velocity at infinity, the initial orbital kinetic energy is still significantly smaller than the total potential energy of the system, hence producing a similar amount of mass lost in the collision, compared to our fiducial model. In other words, our setup is reasonably approximated by a free-fall parabolic encounter, unless a very large velocity is imparted to the impactor (which seems very unlikely, for these stellar masses, in any realistic stellar environment).

In model {\bf primTAMS}, the amount of mass loss is slightly lower ($\approx10$ \%), compared to our standard model. The small difference can be understood by noticing that in this case the primary star is slightly more compact than the one in the standard model, having an initial radius of $\approx 45.6$ R$_\odot$.

\begin{figure}
\includegraphics[scale=0.55]{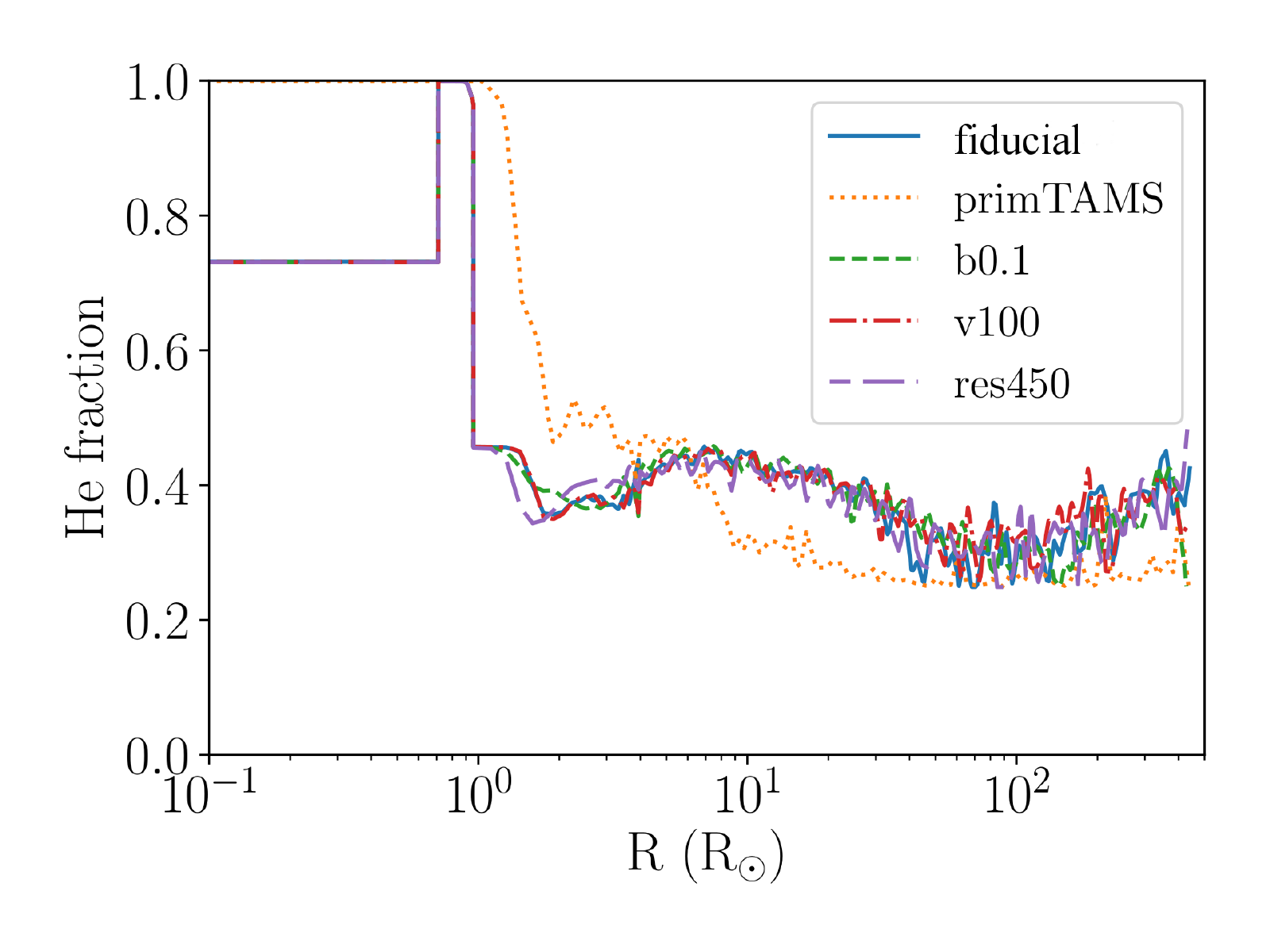}
\caption{Final He profiles of  the additional models described in this appendix, compared to our fiducial model.}\label{He_set}
\end{figure}

Figure \ref{He_set} shows the helium abundance profile of all our models. Also in this case, there is no significant difference between the He profile of  models {\bf res450}, {\bf b0.1}, {\bf v100} and that of our fiducial model. Model {\bf primTAMS} has a different He profile, since the primary star is a terminal-age MS star, hence it has different abundance profiles, compared to those of the other models. Nonetheless, as in the case of our fiducial model, the secondary star gets disrupted at a few solar radii from the center of the primary and deposits its material in the vicinity of the original core. This results in a He profile of the collision product that roughly follows the profile of the primary up to $1-2$~R$_\odot$. 
Going further out, it shows a He fraction that is larger than the original one up to $7-8$~R$_\odot$, roughly matching the one of the core of the secondary ($\approx 0.45$, see Fig. \ref{ics}),  and gets to $\approx 0.25$ in the outer layers, as a result of the chemical abundance of the envelopes of both the primary and secondary star.
 
A further exploration of the parameter space, while challenging with current numerical techniques, is essential to have a more general understanding of massive star--star collisions, and will be the focus of future detailed investigations.



\bsp	
\label{lastpage}
\end{document}